\documentclass[aps,prl,twocolumn]{revtex4}
\makeatletter

\newcommand{\Rmnum}[1]{\expandafter\@slowromancap\romannumeral #1@}
\makeatother
\usepackage{graphicx}
\usepackage{dcolumn}
\usepackage{bm}
\usepackage{CJK}
\usepackage{color}
\usepackage{amsfonts}
\usepackage{psfrag}
\usepackage{wrapfig}
\usepackage{subfigure}
\usepackage{makeidx}
\usepackage{multirow}
\usepackage{epsf}
\usepackage[colorlinks,linkcolor=blue,anchorcolor=red,citecolor=blue,urlcolor=blue]{hyperref}
\usepackage{amsmath}
\usepackage{cases}
\usepackage{bm}
\usepackage[dvipsnames]{xcolor}

\begin{document}
\title{Experimental observation of recurrence and spectral asymmetry of the two-component Akhmediev breathers in a single mode optical fibre}
\author{Chong Liu$^{1,2,3}$}
\author{Le Li$^{1}$}
\author{Shao-Chun Chen$^{1}$}
\author{Xiankun Yao$^{1,2}$}
\author{Wen-Li Yang$^{1,2,3,4}$}
\author{Nail Akhmediev$^5$}
\address{$^1$School of Physics, Northwest University, Xi'an 710127, China}
\address{$^2$Shaanxi Key Laboratory for Theoretical Physics Frontiers, Xi'an 710127, China}
\address{$^3$Peng Huanwu Center for Fundamental Theory, Xi'an 710127, China}
\address{$^4$Institute of Modern Physics, Northwest University, Xi'an 710127, China}
\address{$^5$Department of Fundamental and Theoretical Physics, Research School of Physics and Engineering, The Australian National University, Canberra, ACT 2600, Australia}

\begin{abstract}
We report the results of experimental studies of recurrent spectral dynamics of the two component Akhmediev breathers (ABs) in a single mode optical fibre. We also provide the theoretical analysis and numerical simulations of the ABs based on the two component Manakov equations that confirm the experimental data.
In particular, we observed spectral asymmetry of fundamental ABs and complex spectral evolution of second-order nondegenerate ABs.
\end{abstract}

\maketitle

Fermi-Pasta-Ulam (FPU) recurrence is a fundamental nonlinear phenomenon of spectral expansion and subsequent contraction back to a single mode \cite{FPU0,FPU1,FPU2,Stegeman,FPU3,FPU4,FPU5}. It has been observed experimentally in hydrodynamics \cite{water0,water1,water2}, magnetic films \cite{magnetic}, and in various optical systems \cite{optics0,SG,optics1,optics2,optics3,optics4,optics5}.
The concept of FPU recurrence has led to the foundation of soliton physics \cite{soliton,Book97,Book2006} followed by the inception of the research direction known as `integrable evolution equations' \cite{Gardner,Lax}. The integrable models, in turn, provided ideal platforms to study the FPU recurrence in terms of exact analytical results. One formal mathematical description of FPU recurrence is given by the spectral analysis of Akhmediev breathers (ABs) \cite{AB1}.  Exact AB solutions provide the possibility to calculate the evolution of the infinite number of modes of physical spectra. These infinitely extended spectra, in analytical form, are beyond the reach of the truncated mode approach \cite{TT}. Recent experiments \cite{SD} confirmed the validity of the AB theory through the observation of more than ten spectral sidebands. The first demonstration of one cycle of FPU recurrence in an optical fibre based on a fundamental AB solution was done in \cite{optics0}. Multiple recurrences involving several cycles of AB dynamics were observed recently in fibre optics \cite{SG,optics3} and in spatial optics \cite{optics4}.
More complex dynamics including higher-order AB splitting \cite{Exp2011-PRL,Exp2022-OL} and collision of ABs \cite{Exp2013-PRX} were also observed in fibre optics.

So far, experimental observations of the FPU recurrence have been limited to the simplest scalar case described by the nonlinear Schr\"odinger equation (NLSE). Moving beyond the scalar NLSE in order to model more general classes of physical systems may reveal more involved features of FPU phenomenon.
One of the integrable models of practical importance is the multicomponent NLSEs  known as `Manakov equations' \cite{MM}. It describes variety of complex nonlinear phenomena in fibre \cite{OF1,OF2,OF3,OF4} and spatial optics \cite{SO1,SO2,Chen}), in Bose-Einstein condensates \cite{BEC}, and in the case of two-directional ocean waves \cite{F}. A number of new theoretical and experimental results have been reported on the rich family of vector breather excitations \cite{VB2016-chapter,VB2017-review,VB2012-prl,VB2014-prl,VRW2022,VMI2013,VMI2015,VRW2016-exp,VRW2018-exp,VB2014,VB2019,VSRB-2024,VBDF2022,VB2022,VB2023,VB2021}.
In particular, recurrence phenomena described by the focusing Manakov equations have been observed in fibre optics \cite{VMI2013}. In this work, the frequencies of the two pumps have been chosen equal. As a result, the observed spectra of the two wave components were identical. Consequently, the sidebands were symmetric around the central mode \cite{VMI2013} making no difference from the scalar AB case. Recurrence involving defocusing Manakov equations has been also detected through observation of vector rogue waves  \cite{VRW2016-exp,VRW2018-exp}.
Despite many efforts, the rich dynamics of the vector AB theory remains largely unexplored. This is related to all multicomponent physical systems mentioned above.

Vector ABs reveal nontrivial properties \cite{VB2014,VB2019,VBDF2022,VB2022,VB2023,VB2021} which
differ them from the relatively simple ABs of the scalar NLSE. The most noticeable
difference is that their spectral sidebands exhibit significant asymmetry around the central mode \cite{VBDF2022,VB2022,VB2023,VB2021}.
This asymmetry has been revealed by calculating the exact analytic spectra of the fundamental AB  \cite{VB2021,LA2021}.
Futhermore, a new family of `nondegenerate ABs' that has no analogue in the scalar NLSE case has been found in \cite{VB2022}.
These solutions are formed by nonlinear superposition of two different ABs, having nevertheless equal growth rates. Importantly, these types of
 ABs can be excited using a wide range of initial conditions \cite{VB2022,VB2023} demonstrating their robustness in vector fields.
Clearly, experimental demonstration of these findings would be a crucial step towards practical application of coupled wave systems.

It is known \cite{VMI2015} that propagation of two orthogonally polarised optical pump waves in a randomly birefringent fibre is described by the Manakov equations. These equations, in dimensionless form, are given by \cite{MM}
\begin{equation}\label{eqM}
\begin{split}
i\frac{\partial u_1}{\partial \xi}+\frac{1}{2}\frac{\partial^2 u_1}{\partial \tau^2}+(|u_1|^2+|u_2|^2)u_1 &=0,\\
i\frac{\partial u_2}{\partial \xi}+\frac{1}{2}\frac{\partial^2 u_2}{\partial \tau^2}+(|u_1|^2+|u_2|^2)u_2 &=0,
\end{split}
\end{equation}
where $u_j(\xi,\tau), j=1,2$, are two orthogonally polarised optical pumps while
$\xi$ and $\tau$ are normalized distance and time, respectively.
The fundamental (first-order) AB solution of (\ref{eqM}) is given by \cite{SM}
\begin{eqnarray}
u_j=u_{0j}\left[\frac{\cosh(\bm{\Gamma}+i\gamma_j)e^{i\eta_{1j}}+\varpi \cos(\bm{\Omega}-i\epsilon_j)e^{i\eta_{2j}}}{\cosh\bm{\Gamma}+\varpi \cos\bm{\Omega}}\right].\label{eqb}
\end{eqnarray}
Here $u_{0j}=a_j \exp \left\{ i \left[{\omega_j}\tau + (a_1^2+a_2^2- \omega_j^2/2) \xi \right] \right\}$ are the two continuous wave pumps with the amplitudes $a_j=a$ and
the frequencies $\omega_j$.  We choose different frequencies $\omega_1=-\omega_2=\Delta\omega/2$ in order to observe nontrivial properties of vector ABs. When $\Delta\omega=0$, solution (\ref{eqb}) reduces to the scalar AB.
The scalar arguments $\bm{\Gamma}$ and $\bm{\Omega}$ are:
\begin{eqnarray}
\bm{\Gamma}=\omega {\chi}_i\bm \xi,~
\bm{\Omega}=\omega \left[\bm \tau+ ({\chi}_r+\frac{1}{2}\omega )\bm \xi \right]+\arg\frac{2{\chi}_i}{2{\chi}_i-i\omega}.
\end{eqnarray}
Here $\bm{\xi}=\xi-\xi_{1}$, $\bm{\tau}=\tau-\tau_{1}$ are shifted spatial and time variables respectively with $\xi_{1}$ and $\tau_{1}$ defining the spatial and temporal position of the breather. Parameter $\omega$ is the initial modulation frequency.
Other notations in (\ref{eqb}) are the same as in Ref. \cite{VB2022}. In particular, one of the important parameters of the AB is its complex eigenvalue ${\chi}$ with its real ${\chi}_r$ and imaginary ${\chi}_i$ parts, given by:
\begin{eqnarray}
{\chi}_{\pm}=\pm(\Delta\omega^2/4-a^2+\omega^2/4-\sqrt{{\nu}})^{1/2}-\omega/2,\label{eqchi}
\end{eqnarray}
where ${\nu}=a^4-a^2\Delta\omega^2+\omega^2\Delta\omega^2/4$.

Equation (\ref{eqb}) describes a three-parameter family of solutions depending on  $a$, $\Delta\omega$, and $\omega$.
It represents the full growth-decay cycle of modulation instability (MI) with the growth rate $G=|\omega {\chi}_i|$. ABs exist when $G\neq0$. The physical spectrum of (\ref{eqb}) is given by
\begin{equation}
A_{j,n}(\xi)=\frac{\omega}{2\pi}\int_0^{2\pi/\omega}u_{j}(\tau,\xi) e^{-in\omega \tau}d \tau. \label{eqs}
\end{equation}
All modes $A_{j,n}~(n=0, \pm1, \pm2$, ...) evolve in $\xi$.
Their exact analytic expressions are given in \cite{VB2021}.
These components describe the full expansion-contraction cycle of the spectrum. As soon as $\Delta\omega\neq0$, the sidebands of the spectra $|A_{j,n\neq0}|$ become asymmetric i.e., $|A_{j,n}|\neq|A_{j,-n}|$.

\begin{figure}[htb]
\centering
\includegraphics[width=85mm]{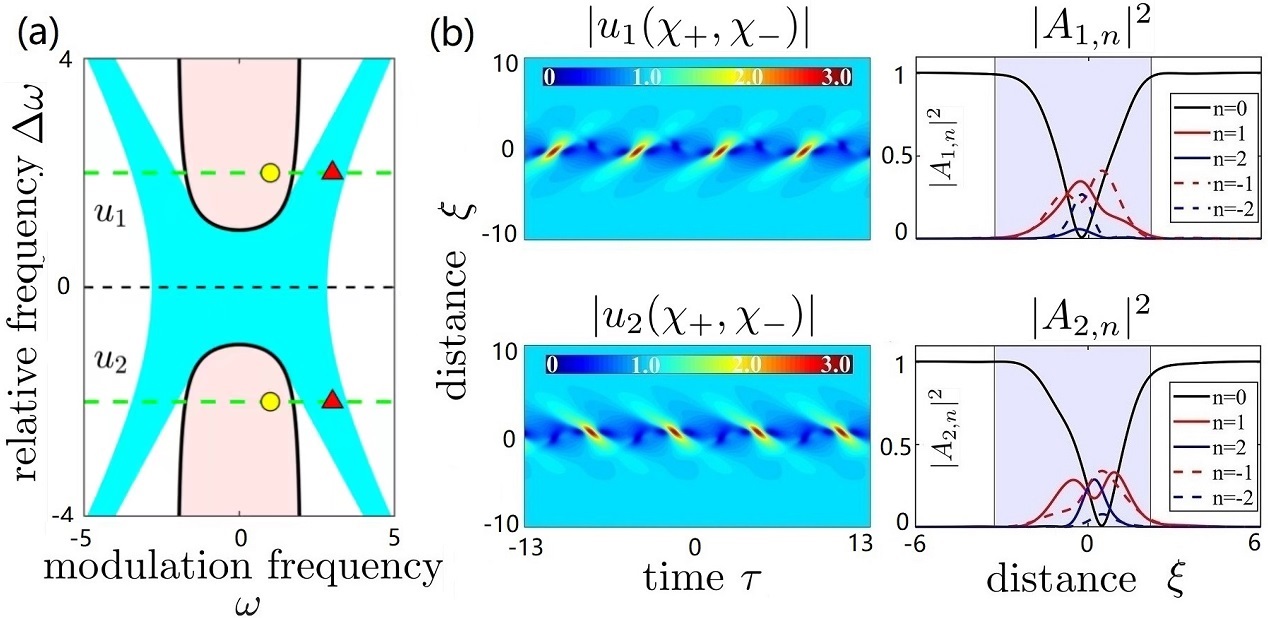}
\caption{
(a) Existence diagram of vector ABs. Pink areas denote the existence regions of the second-order nondegenerate ABs with $\omega^2\leq\omega_c^2$.
The circles and triangles correspond to the ABs with parameters $(\omega,\Delta\omega)=(1,2)$ and $(\omega,\Delta\omega)=(3,2)$, respectively.
(b) (left panels) The amplitude profiles $|u_j({\chi}_{+}, {\chi}_{-})|$ and (right panels) the spectral intensities $|A_{j,n}|^2$ of nondegenerate ABs corresponding to the circles in (a). Parameter $a=1$.}\label{f-AB-1}
\end{figure}

Figure \ref{f-AB-1} (a) shows the AB existence diagram defined by $G=|\omega {\chi}_i|\neq0$ on the ($\omega,\Delta\omega$) plane. A critical frequency is given by $\omega_c^2=4a^2-4a^4/\Delta\omega^2$ (solid lines) when $\nu=0$. Nondegnerate ABs \cite{VB2022} exist in the pink areas where $\omega^2\leq\omega_c^2$. There is only one AB in the cyan areas.

In the pink region, ${\chi}_{+,i}=-{\chi}_{-,i}$, but ${\chi}_{+,r}\neq{\chi}_{-,r}$.
This means that there are two different AB solutions, $u_j({\chi}_{+})\neq u_j({\chi}_{-})$, for given initial parameters despite the fact that they have the equal values of the growth rate $G({\chi}_{+})=G({\chi}_{-})$.
Their nonlinear superposition results in the second-order `nondegenerate ABs', $u_j({\chi}_{+},{\chi}_{-})$. For a given set of external parameters, the dynamics of nondegenerate ABs depends on the relative separations in time and space of the individual ABs, i.e. $(\tau_1,\xi_1,\tau_2,\xi_2)$.
To give an example,
Figure \ref{f-AB-1} (b) shows
the amplitude profiles $|u_j|$ and the evolution of the spectra $|A_{j,n}|^2$ of the nondegenerate ABs, when $(\tau_1,\xi_1,\tau_2,\xi_2)=(1.28, 0.17,-1.60,-0.78)$.
The amplitude profile shows complex periodic structure (four-petal pattern) in contrast to the fundamental AB with the same period $2\pi/\omega$ in $\tau$.
The sidebands are asymmetric $|A_{j,n}|\neq|A_{j,-n}|$ which is one of the main points of our present study.

\begin{figure}[htb]
\centering
\includegraphics[width=85mm]{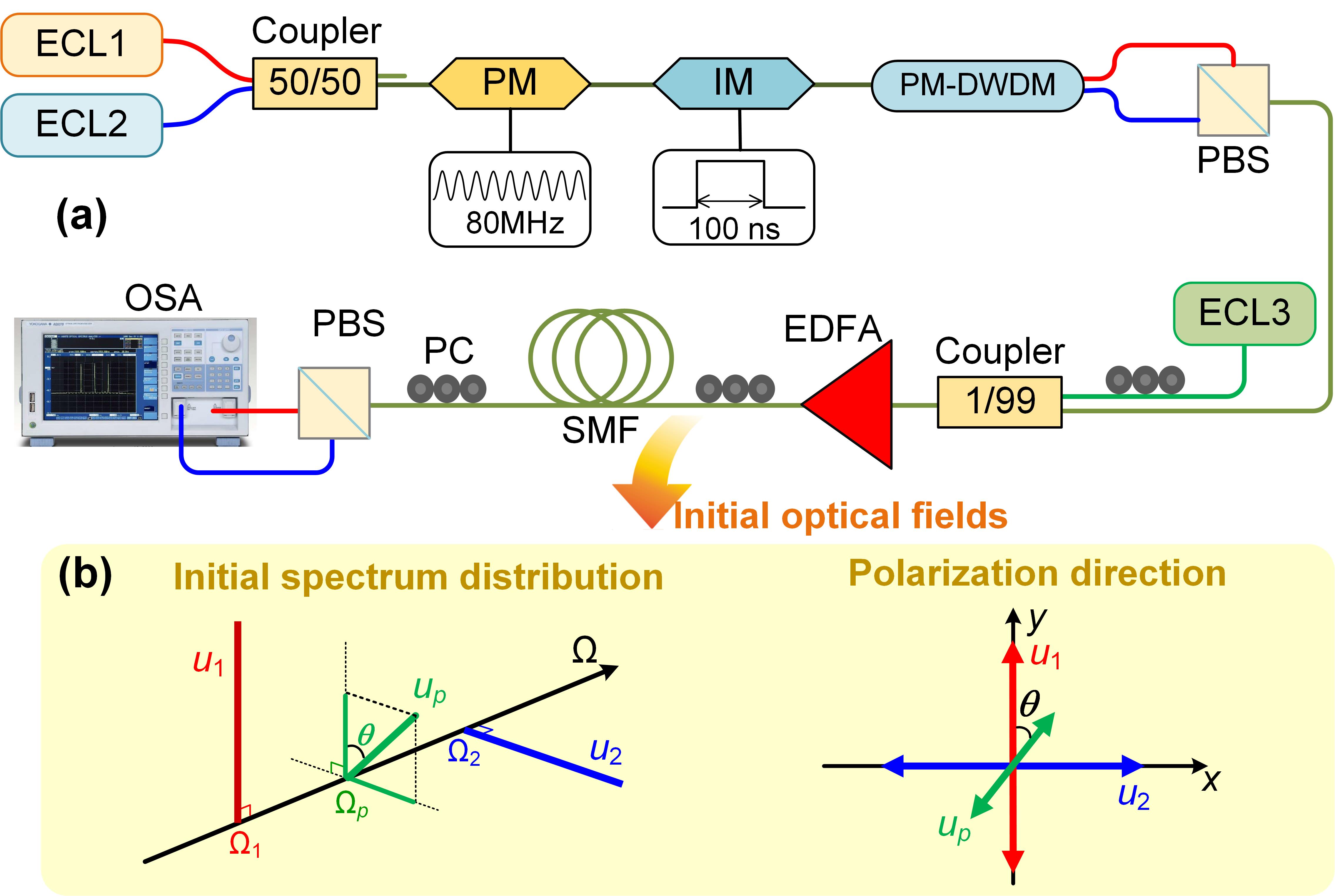}
\caption{Schematic diagram of the experimental setup. Here, ECL1 and ECL2 are the two external laser diodes; PM is phase modulator; IM is the intensity modulator;
PM-DWDM is the polarization-maintaining dense wavelength division multiplexer; PBS is the polarization beam splitter; PC is polarisation controller;
EDFA is erbium-doped fiber amplifier; SMF is a single-mode fibre; OSA is the optical spectrum analyser. Red and blue lines show the propagation of two orthogonally polarised pump waves while the green lines show the propagation of probe wave. The probe light is polarised at an angle $\theta$ to the component $U_1$.
}\label{f-exp-setup}
\end{figure}

Our experimental setup is shown in Fig. \ref{f-exp-setup}.
Here, the two diode lasers (ECL1, ECL2) generate the pump waves $U_j$ at frequencies $\Omega_j$ with a frequency difference $\Delta\Omega=2\pi\times100~\textmd{GHz}$.
An 80~MHz RF signal drives the phase modulator PM to achieve higher peak pumping power and to suppress stimulated Brillouin scattering.
The intensity modulator IM is driven by waveform transmitter
 which generates a 100~ns flat-top pulse with a 1:10 duty cycle to provide a quasi-continuous wave background. The wavelength division multiplexer DWDM separates the two frequencies of light, which are subsequently combined into orthogonally polarised waves using a polarization beam synthesizer PBS.

Each of the two components of the pump wave, $U_1$ and $U_2$, is amplified up to the value $P_{j}=P_0=1.6\textmd{W}$ in the erbium-doped fibre EDFA. A weak probe wave that corresponds to the wavelength 1560 nm is generated by the laser diode ECL3 at frequency $\Omega_p$.
The power of the probe wave is relatively small, namely, $P_{p}$/$P_0=0.0225(\ll1)$.
It is polarised at an angle $\theta$ to the pump component $U_1$ after passing
the polarisation controller.

The modulated input to the single-mode fibre, SMF, is given by  \cite{SM}
\begin{eqnarray}\label{eqin0}
\begin{split}
U_1(0,t)=\sqrt{P_0}e^{i(\Omega_{1} t+\phi_1)}\left[1+a^{(1)}_{\textrm{mod}}e^{i(\Delta\phi_1 \pm\Omega^{(1)}_{\textrm{mod}} t)}\right],\\
U_2(0,t)=\sqrt{P_0}e^{i(\Omega_{2} t+\phi_2)}\left[1+a^{(2)}_{\textrm{mod}}e^{i(\Delta\phi_2 \pm\Omega^{(2)}_{\textrm{mod}} t)}\right],
\end{split}
\end{eqnarray}
where the two modulation amplitudes are $a^{(1)}_{\textrm{mod}}=\sqrt{P_p/P_0}\cos\theta$, $a^{(2)}_{\textrm{mod}}=\sqrt{P_p/P_0}\sin\theta$, with
$\theta\in[0,\pi/2]$, and
the modulation frequencies are $\Omega^{(1)}_{\textrm{mod}}=\Omega_{p}-\Omega_{1}$, $\Omega^{(2)}_{\textrm{mod}}=\Omega_{2}-\Omega_{p}$.
The phase differences are $\Delta\phi_1=\phi_p-\phi_1$, $\Delta\phi_2=\phi_p-\phi_2$, where $\phi_1$, $\phi_2$ and $\phi_p$ are random phases induced by the laser diodes.

The nonlinear stage of propagation induced by the initial condition (\ref{eqin0})
starts in a single-mode fibre, SMF.
The group velocity dispersion of the SMF is $\beta_2=-22 \textmd{ps}^2\textmd{km}^{-1}$, the nonlinear coefficient $\gamma=1.1 \textmd{W}^{-1}\textmd{km}^{-1}$, and the fiber loss $\alpha=0.2 \textmd{dB} \textmd{km}^{-1}$.
The spectral output of the SMF  is measured by an optical spectrum analyser, OSA.
The two wave components $U_j$ are separated by a polarising beam splitter, PBS.
The spectral evolution is studied using different lengths of the same fiber.

Initial condition (\ref{eqin0}) is the linear MI approximation rather than the exact
expression that follows from the vector AB solution.
Changing $\theta$, we can adjust the initial modulation ratio of the two components. When $\theta=\pi/4$, the modulation amplitudes are equal,
$a^{(1)}_{\textrm{mod}}=a^{(2)}_{\textrm{mod}}$. Otherwise, $a^{(1)}_{\textrm{mod}}\neq a^{(2)}_{\textrm{mod}}$. Below, we consider the case $\theta=0$.
The case $\theta=\pi/4$ is analysed in Supplementary Material \cite{SM}.

When $\theta=0$, $a^{(2)}_{\textrm{mod}}=0$. Then, only the first component $U_1$ is modulated:
\begin{eqnarray}\label{eqin2}
\begin{split}
U_1(0,t)&=\sqrt{P_0}e^{i(\Omega_{1} t+\phi_1)}\left[1+a^{(1)}_{\textrm{mod}}e^{i(\Delta\phi_1+\Omega^{(1)}_{\textrm{mod}} t)}\right],\\
U_2(0,t)&=\sqrt{P_0}e^{i(\Omega_{2} t+\phi_2)},
\end{split}
\end{eqnarray}
 In this case, none of the phases $\phi_j$, or $\Delta\phi_1$ influences the dynamics of the AB spectra. The remaining variable parameter is $\Omega_p$. Its alteration changes the AB spectral dynamics from nondegenerate AB pattern to four-petal structures.
In the experiments, we set $\Omega^{(1)}_{\textrm{mod}}/\Delta\Omega=\omega/\Delta\omega$ and choose the initial modulation frequency $\Omega^{(1)}_{\textrm{mod}}=2\pi\times50~\textmd{GHz}$.

\begin{figure}[htb]
\centering
\includegraphics[width=85mm]{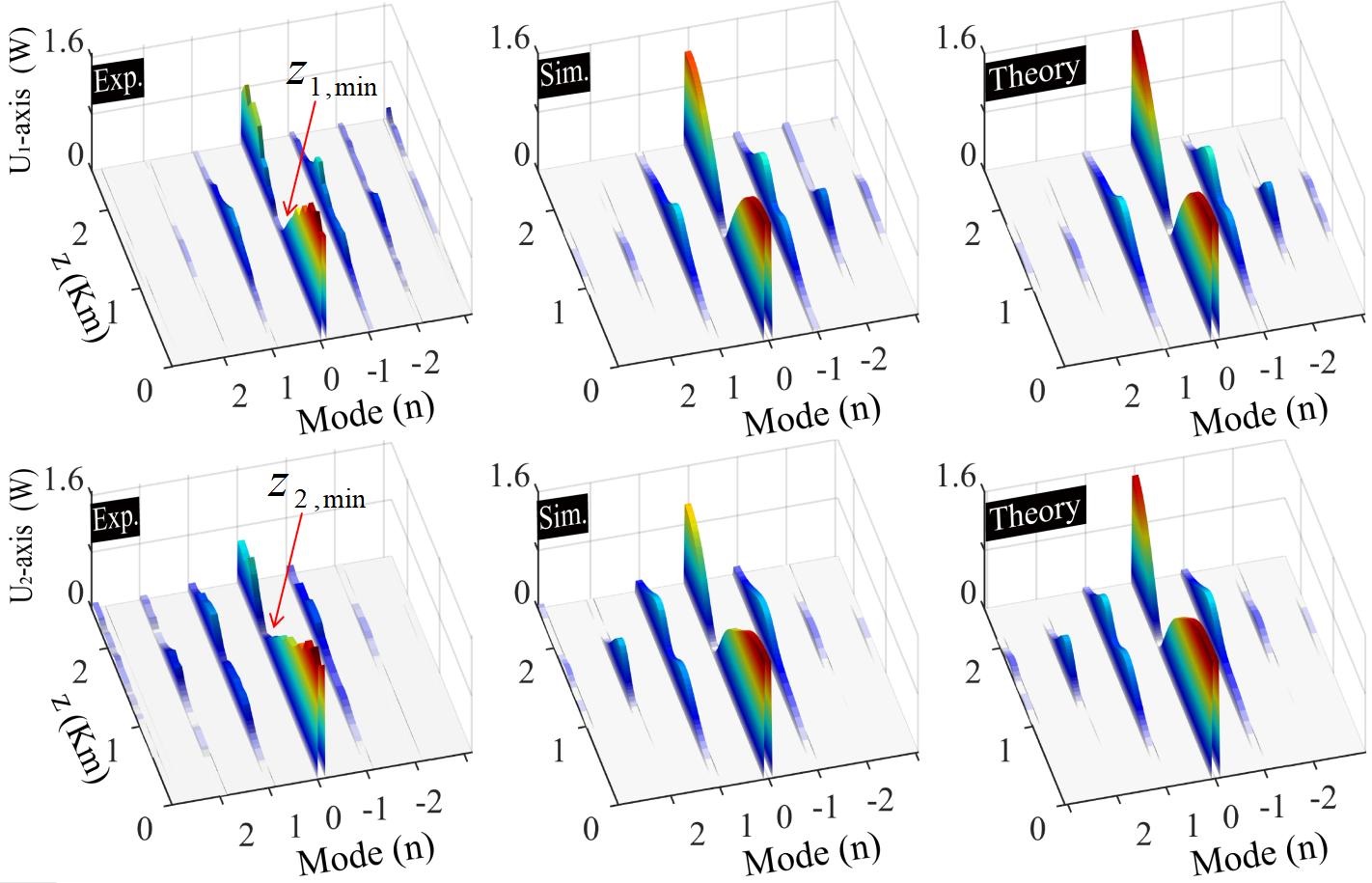}
\caption{Evolution of the spectral modes of the second-order nondegenerate ABs excited using the initial condition (\ref{eqin2}).
The upper row panels correspond to the first component $U_1$, the lower row shows the spectral evolution of the second component $U_2$. The left hand side panels are the experimental data. The middle column panels -- numerical simulations. The right hand side panels -- theoretical results.
}\label{f-exp-0}
\end{figure}

Figure \ref{f-exp-0} shows spectral mode evolution of nondegenerate ABs. Left hand side panels correspond to the experimental data, the central panels show the results of numerical simulations while the curves of the right hand side panels are calculated using the exact theory.
Parameters $(\omega, \Delta\omega)=(1,2)$ are the same as in Fig.
\ref{f-AB-1} (a).

Initially, at $z=0$, almost all energy ($P_{j}=P_0=1.6\textmd{W}$) is concentrated in the pumps.
With increasing $z$, the pump energy is dispersed between the sidebands.
Minimal energy of the pump is reached at $z=z_{j,\textmd{min}}$. In the case of nondegenerate ABs, the points of minima are different for the two components.
The difference $\Delta z=z_{1,\textmd{min}}-z_{2,\textmd{min}}>0$. In other cases,
$\Delta z=0$. Beyond the points of minima, spectral energy returns back to the pump mode as predicted by the AB theory. This can be considered as the FPU recurrence in vector case.

Numerical simulations starting with the initial condition (\ref{eqin2}) are performed by solving the Manakov equations using the split-step Fourier method. Intensities of the spectral components of the ABs directly follow from these simulations. Theoretical curves are calculated using the exact solutions presented above.
As can be seen from Fig. \ref{f-exp-0}, both numerical simulations and theory confirm our experimental observations.

\begin{figure}[htb]
\centering
\includegraphics[width=85mm]{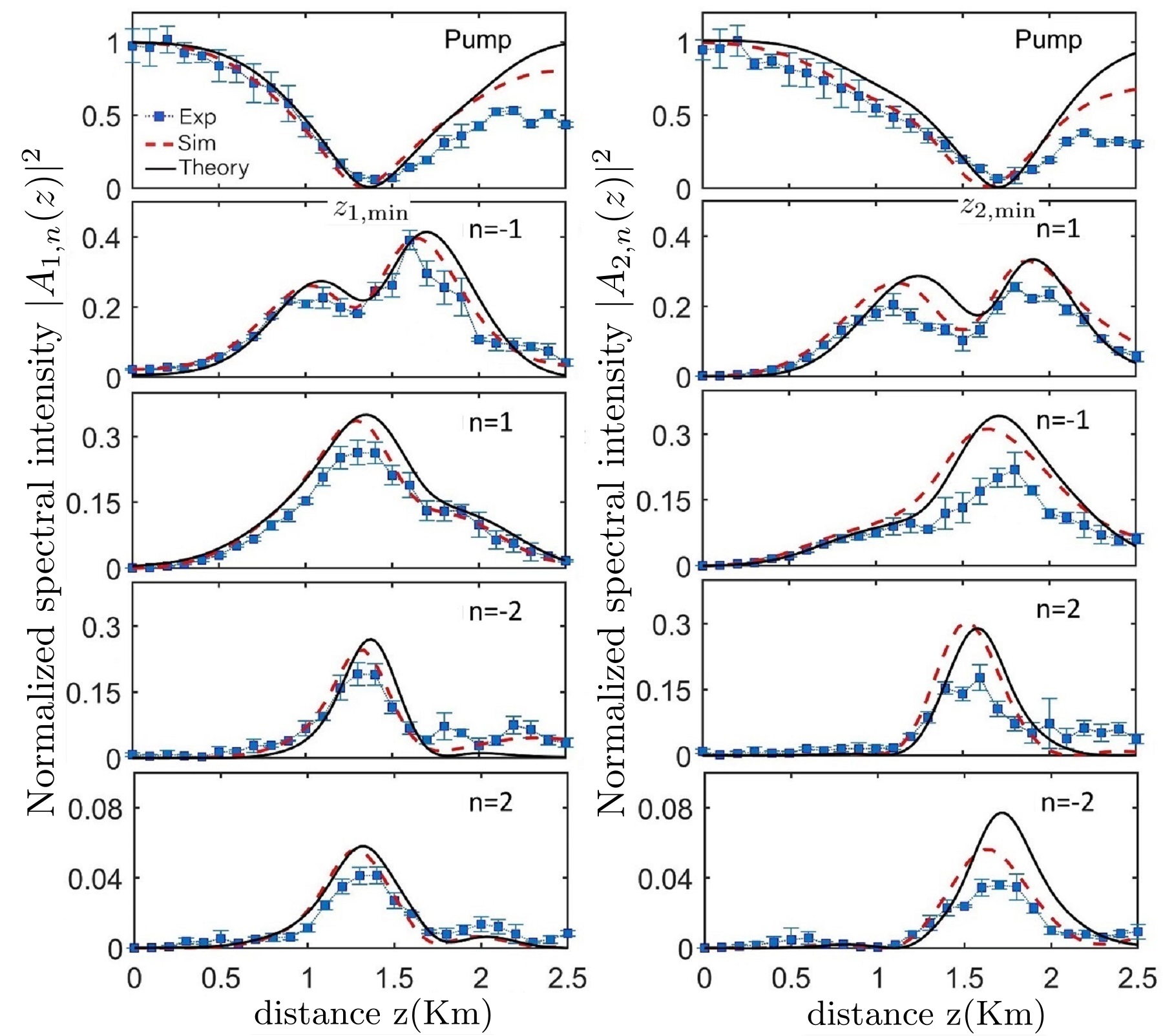}
\caption{Evolution of the pump, $|A_{j,0}(z)|^2$, (upper panels) and the lowest order sidebands $|A_{j,n}(z)|^2$ of nondegenerate ABs for $n=\pm1,\pm2$. Left hand side panels show the evolution of the first component ($j=1$), the right hand side panels show the evolution of the second component ($j=2$). Blue rectangles show the experimental data, red dashed curves correspond to numerical simulations, and the black solid curves are theoretical results.
}\label{f-exp-0-1}
\end{figure}

More accurate depiction of the curves for the pump ($n=0$) and the lowest order ($n=\pm1,\pm2$) spectral mode evolution that includes experimental error bars is presented in Fig. \ref{f-exp-0-1}. The spectral curves for the first and the second components are distinctly different. The error is calculated as the standard deviation of multiple (five) measurements. Within this error, the experimental spectra are in good agreement with the analytical results and numerical simulations. The difference is caused by the inaccuracy of initial conditions in the experiment and the simulations.

\begin{figure}[htb]
\centering
\includegraphics[width=85mm]{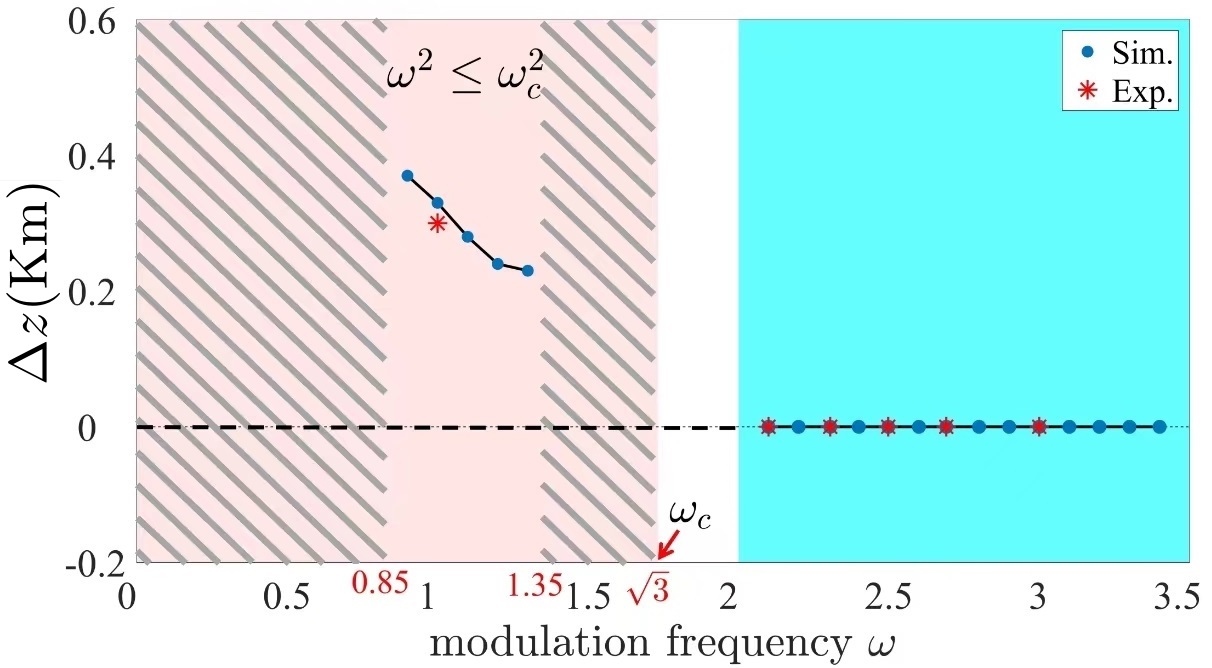}
\caption{
Difference $\Delta z$ between the locations of minimal pump energy  versus modulation frequency $\omega=\Delta\omega\Omega^{(1)}_{\textrm{mod}}/\Delta\Omega$. Pink and cyan areas correspond to the regions of the same color shown in Fig. \ref{f-AB-1}(a) with $\Delta\omega=2$. Blue dots show the results of numerical simulations. Red stars follow from the experimental data.}
\label{f-exp-0-2}
\end{figure}

The asymmetry is observed not only in the shapes of the spectral curves but also in the different locations of the points of minima reached by the two pumps. This can be seen from comparison of the two upper panels in Fig. \ref{f-exp-0-1}. The distance between the points of minima $\Delta z=z_{1,\textmd{min}}-z_{2,\textmd{min}}$ depends on modulation frequency $\omega$. This dependence
is shown in Fig. \ref{f-exp-0-2}.
Nondegenerate ABs in this plot exist in the pink region below the critical frequency,  $0<\omega<\omega_c=\sqrt{3}$. In the pure pink region of nondegenerate ABs located within the limits, $0.85\leq\omega\leq1.35$, $\Delta z$ decreases with increasing $\omega$. In the two shaded pink area, the effects of higher-order MI result in significantly more complicated spectra \cite{VB2023}. We omit their analysis here.
The spectral region $\sqrt{3}<\omega<2$ is stable relative to the frequency modulations. Thus, there are no AB solutions in this (white) region.
Above the frequency $\omega>2$ (cyan area), $\Delta z=0$. This is the area of fundamental ABs. Numerical simulations (blue circles) and the six experimental points (red stars) confirm this fact.

In conclusion, we report experimental studies of asymmetric spectral recurrent dynamics of vector ABs in a single mode optical fibre. Our experimental results are confirmed by numerical simulations and theoretical analysis of
Manakov equations.
This work is a significant new step towards understanding of nontrivial spectral dynamics of ABs in the vector case.
Because of the universal applicability of Manakov equations in physics of complex systems, our results may have a notable impact on further developments
in several areas of physics including quantum atom physics and water waves.


\end{document}